\documentclass[epj]{webofc}
\usepackage[varg]{txfonts}   
\newcommand{\bfk}{\mathbf k}
\newcommand{\bfr}{\mathbf r}

\DeclareMathOperator{\erf}{Erf}

\wocname{EPJ Web of Conferences}
\woctitle{INPC 2013}
\begin{document}
\title{New density-independent interactions for nuclear structure calculations}
%
%

\author{K. Bennaceur\inst{1} \and
        J. Dobaczewski\inst{2,3} \and
        F. Raimondi\inst{4}
}

\institute{
{Universit\'e de Lyon, F-69003 Lyon, France;
Institut de Physique Nucl\'eaire de Lyon, CNRS/IN2P3, Universit\'e Lyon 1,
F-69622 Villeurbanne Cedex, France}
\and
{Institute of Theoretical Physics, Faculty of Physics, University of Warsaw, ul. Ho{\.z}a 69, PL-00681 Warsaw, Poland}
\and
{Department of Physics, PO Box 35 (YFL), FI-40014
University of Jyv{\"a}skyl{\"a}, Finland}
\and
{TRIUMF, 4004 Wesbrook Mall, Vancouver, British Columbia V6T2A3, Canada}
          }

\abstract{%
We present a new two-body finite-range and momentum-dependent but density-independent
effective interaction, which can be interpreted as a regularized
zero-range force. We show that no three-body or density-dependent
terms are needed for a correct description of saturation properties in
infinite matter, that is, on the level of low-energy density
functional, the physical three-body effects can be efficiently
absorbed in effective two-body terms. The new interaction gives a
satisfying equation of state of nuclear matter and opens up
extremely interesting perspectives for the mean-field and
beyond-mean-field descriptions of atomic nuclei.
}
\maketitle
%
The two most widely used effective forces for nonrelativistic
energy-density-functional
calculations, the Gogny~\cite{Decharge80,Berger91} and the Skyrme
interaction~\cite{Skyrme56-58}, have in common a two-body density
dependent term.
For the case of the Skyrme force, the original three-body contact force of the form $t_3\delta(\bfr_1-\bfr_2)\delta(\bfr_1-\bfr_3)$~\cite{Skyrme56-58} was in fact replaced
by an isoscalar density-dependent
two-body contact force, $\tfrac{1}{6}\,
t_3\left(1+x_3\hat P^\sigma\right)\rho_0^\alpha\left(\tfrac{\bfr_1+\bfr_2}{2}\right)
\delta(\bfr_1-\bfr_2)$, due to the appearance of
spin instabilities~\cite{Chang75,Backman75,Waroquier76}. Later, it was
employed
as a convenient way of simulating the density dependence of the effective
interaction rather than a three-body force~\cite{Beiner75}.

For Skyrme and Gogny effective interactions, the density
dependent term appears to play a determinant role, especially
in generating the mechanism of saturation. The non-integer powers
of the density seem to be mandatory if one wants to obtain
acceptable values for incompressibility and isoscalar effective mass.
These considerations concern only the description of nuclei at the
mean-field level. However, several years ago it was identified that
effective interactions that depend on non-integer powers of
density do not allow for beyond-mean-field calculations that
use standard techniques of symmetry restoration or
Generator Coordinate Method~\cite{[Dob07d],Duguet08}.

This observation triggered efforts to define a new generation of
effective interactions without density dependence. A promising way is
to consider two- and three-body momentum dependent zero-range terms
along with a four-body momentum independent one~\cite{Sadoudi11}.
Another way, discussed in this study, consists in using a form very
similar to the two-body part of the Skyrme interaction and replacing
the $\delta$ Dirac functions by finite-range form factors. This
corresponds to the next-to-leading-order expansion of the effective
interaction introduced in Ref.~\cite{Dobacweski12}.

Below we show that the use
of a finite-range mo\-men\-tum-\-de\-pen\-dent, but density-independent,
two-body interaction allows us to describe most of the medium effects in a
realistic way.

We use the notation $x\equiv(\bfr,\sigma,q)$ where $\sigma$ and $q$ are
the spin and isospin projections. For a general nonlocal
two-body effective interaction $v$, the mean-field average value of the potential
energy can be written as
%
$\langle V\rangle=\frac{1}{2}\int\!
\mathrm dx_1\,\mathrm dx_2\,\mathrm dx_3\,\mathrm dx_4\,
v(x_1,x_2;x_3,x_4)
\left[\rho(x_3,x_1)\rho(x_4,x_2)-\rho(x_4,x_1)\rho(x_3,x_2)\right]$ ,
%
where $\rho$ is the nonlocal one-body density.
Following Ref.~\cite{Dobacweski12},
we choose the interaction as:
\vspace*{-0.4cm}
\begin{align}
\label{eq:bdr}
v&= \sum_{k=0}^2
\left[T_k^{(1)}\hat\delta^{\sigma q}+T_k^{(2)}\hat\delta^q\hat P^\sigma
-T_k^{(3)}\hat\delta^\sigma\hat P^q-T_k^{(4)}\hat P^\sigma \hat P^q\right]
\hat{O}_k(\bfk_{12}^*,\bfk_{34})\delta(\bfr_1-\bfr_3)\delta(\bfr_2-\bfr_4)
g_a(\bfr_2-\bfr_1) , \\[-10mm] \nonumber
\end{align}
where $\hat P^\sigma$ and $\hat P^q$ denote the standard spin and isospin
exchange operators, respectively,
$\hat \delta^\sigma$, $\hat \delta^q$, and
$\hat \delta^{\sigma q}$ are the identity operators in the spin and/or
isospin spaces, and
%
$\hat{O}_0 = 1$,
$\hat{O}_1 = \tfrac{1}{2}\left[\bfk_{12}^{*2}+\bfk_{34}^2\right]$,
$\hat{O}_2 = \bfk_{12}^*\cdot\bfk_{34}$, and
$g_a(\bfr)=e^{-\bfr^2/a^2}/(a\sqrt{\pi})^3$.
Here, ${\bfk}_{ij}$ represents the relative momentum operator between
particles $i$ and $j$ and the regularized $\delta$ function $g_a(\bfr)$
defines the profile of the interaction, with $a$ representing the
regularization scale and range of the interaction. This effective
interaction depends on 13 parameters, that is, on 12 parameters
$T_k^{(i)}$, for $k=0,1,2$ and $i=1,2,3,4$, which control the
strengths in each channel, and on the range $a$. As usual,
integration by parts of matrix elements transforms the nonlocal
interaction (\ref{eq:bdr}) into a local finite-range
momentum-dependent pseudopotential.
Although a construction of the analogous finite-range spin-orbit force
poses no difficulty, in the present implementation we use the standard
zero-range form of it, with one strength parameter $W_0$.

By using methods of symbolic programming, one can derive
the nuclear-matter characteristics of interaction (\ref{eq:bdr}).
For that, we introduce the following auxiliary functions:
\vspace*{-0.4cm}
\begin{align}
F_0(\xi)&=\frac{12}{\xi^3}\left[\frac{1-e^{-\xi^2}}{\xi^3}-\frac{3-e^{-\xi^2}}{2\xi}
+\frac{\sqrt{\pi}}{2}\erf \xi\right]\,, \\
F_1(\xi)&=\frac{48}{\xi^8}\left(1-e^{-\xi^2}\right)
+\left(\frac{12}{\xi^6}+\frac{6}{\xi^4}\right)\left(e^{-\xi^2}-5\right)
+\left(1+\frac{5}{\xi^2}\right)\frac{6\sqrt{\pi}}{\xi^3}\erf \xi\,, \\
F_2(\xi)&=\frac{720}{\xi^{10}}\left(e^{-\xi^2}-1\right)
+\frac{120}{\xi^8}\left(e^{-\xi^2}+5\right)
+\frac{60}{\xi^6}\left(e^{-\xi^2}-5\right)
+\frac{60\sqrt{\pi}}{\xi^5}\erf \xi\,, \\
%
G_0(\xi)&=\frac{12}{\xi^6}\left(e^{-\xi^2}-1\right)
+\frac{6}{\xi^4}\left(e^{-\xi^2}+1\right)\,, \\
G_1(\xi)&=\frac{12}{\xi^6}\left(1-e^{-\xi^2}\right)-\frac{12}{\xi^4}
-\frac{3}{\xi^2}\left(1+e^{-\xi^2}\right)+\frac{6\sqrt{\pi}}{\xi^3}\erf \xi\,, \\
G_2(\xi)&=\frac{60}{\xi^8}\left(e^{-\xi^2}-1\right)
+\frac{30}{\xi^6}\left(3e^{-\xi^2}-1\right)
+\frac{15}{\xi^4}\left(e^{-\xi^2}-1\right)
+\frac{30\sqrt{\pi}}{\xi^5}\erf \xi\,, \\[-8.8mm] \nonumber
\end{align}
and
$H_0(\xi)=\left(1-e^{-\xi^2}\right)/\xi^2$,
$H_1(\xi)=G_0(\xi)$ .
In the following, we use these functions at $\xi=k_F a$,
which fixes their dependence on the Fermi momentum $k_F$ and range $a$.
With definitions
$A^{\rho_0}_i = \phantom{-}\tfrac{1}{2}T_1^{(i)}+\tfrac{1}{4}T_2^{(i)}-\tfrac{1}{4}T_3^{(i)}-\tfrac{1}{8}T_4^{(i)}$,
$B^{\rho_0}_i =          - \tfrac{1}{8}T_1^{(i)}-\tfrac{1}{4}T_2^{(i)}+\tfrac{1}{4}T_3^{(i)}+\tfrac{1}{2}T_4^{(i)}$,
$A^{\rho_1}_i =                                                       -\tfrac{1}{4}T_3^{(i)}-\tfrac{1}{8}T_4^{(i)}$, and
$B^{\rho_1}_i =          - \tfrac{1}{8}T_1^{(i)}-\tfrac{1}{4}T_2^{(i)}                                            $
we then have the
energy per particle, $\tfrac{E}{A}$,
incompressibility, $K_\infty$,
effective mass, $m^*$,
symmetry energy, $J$,
slope of the symmetry energy, $L$, and
curvature of the symmetry energy, $K_\mathrm{sym}$, given as:
\vspace*{-0.4cm}
\begin{align}
\frac{E}{A}=&\frac{\hbar^2}{2m}\frac{\tau_0}{\rho_0}
+\left[A_0^{\rho_0}+B_0^{\rho_0}F_0(\xi)\right]\rho_0
+\frac{1}{2}\left(A_1^{\rho_0}+A_2^{\rho_0}\right)\tau_0
+\frac{1}{2}\left(B_1^{\rho_0}-B_2^{\rho_0}\right)
\left[F_1(\xi)-\frac{\xi^2}{10}
F_2(\xi)\right]\tau_0 \,,\\[-0.7mm] \nonumber
K_\infty=&-2\frac{\hbar^2}{2m}\frac{\tau_0}{\rho_0}
+B_0^{\rho_0}\left[4\xi F_0'(\xi)+\xi^2F_0''(\xi)\right]\rho_0
+5\left(A_1^{\rho_0}+A_2^{\rho_0}\right)\tau_0\\[-0.7mm]
+&\frac{1}{2}\left(B_1^{\rho_0}-B_2^{\rho_0}\right)
\left[
10F_1(\xi)+8\xi F_1'(\xi)+\xi^2 F_1''(\xi)
-\frac{14\xi^2}{5}F_2(\xi)-\frac{6\xi^3}{5}F_2'(\xi)
-\frac{\xi^4}{10}F_2''(\xi)\right] \tau_0 \,, \\[-0.7mm]
\frac{\hbar^2}{2m^*}=&\frac{\hbar^2}{2m}
-\frac{1}{2}B_0^{\rho_0}a^2\rho_0G_0(\xi)
+\frac{1}{2}\left(A_1^{\rho_0}+A_2^{\rho_0}\right)\rho_0
+\frac{1}{2}\left(B_1^{\rho_0}-B_2^{\rho_0}\right)\rho_0
\left[G_1(\xi)-\frac{\xi^2}{5}G_2(\xi)\right]  \,,  \\[-17mm]    \nonumber
\end{align}
\begin{align}
J=&\frac{5}{9}\frac{\hbar^2}{2m}\frac{\tau_0}{\rho_0}
-B_0^{\rho_0}\frac{\xi^2}{6}G_0(\xi)\rho_0
+\frac{5}{18}\left(A_1^{\rho_0}+A_2^{\rho_0}\right)\tau_0
+\frac{5}{18}\left(B_1^{\rho_0}-B_2^{\rho_0}\right)
\left[G_1(\xi)-\frac{\xi^2}{5}G_2(\xi)\right]\tau_0 \nonumber \\[-1.1mm]
&+\left[A_0^{\rho_1}+B_0^{\rho_1}H_0(\xi)\right]\rho_0
+\frac{5}{6}\left(A_1^{\rho_1}+A_2^{\rho_1}\right)\tau_0
+\frac{5}{6}\left(B_1^{\rho_1}-B_2^{\rho_1}\right)
\left[H_0(\xi)-\frac{\xi^2}{6}H_1(\xi)\right]\tau_0
   \,, \\[-1.1mm]
L=&\frac{10}{9}\frac{\hbar^2}{2m}\frac{\tau_0}{\rho_0}
-B_0^{\rho_0}
\left[\frac{5\xi^2}{6}G_0(\xi)+\frac{\xi^3}{6}G_0'(\xi)\right]\rho_0
+\frac{25}{18}\left(A_1^{\rho_0}+A_2^{\rho_0}\right)\tau_0
+3A_0^{\rho_1}
\rho_0  \nonumber
 \\[-1.1mm]
+&\frac{1}{18}\left(B_1^{\rho_0}-B_2^{\rho_0}\right)
\left[25G_1(\xi)-7\xi^2G_2(\xi)+5\xi G_1'(\xi)-\xi^3G_2'(\xi)
\right]\tau_0
+3B_0^{\rho_1}\left[H_0(\xi)+\frac{\xi}{3} H_0'(\xi)\right]\rho_0
  \nonumber \\[-1.1mm]
+&\frac{25}{6}\left(A_1^{\rho_1}+A_2^{\rho_1}\right)\tau_0
+\frac{25}{6}\left(B_1^{\rho_1}-B_2^{\rho_1}\right)
\left[H_0(\xi)+\frac{\xi}{5}H_0'(\xi)
-\frac{7\xi^2}{30}H_1(\xi)-\frac{\xi^3}{30}H_1'(\xi)\right]\tau_0
    \,, \\[-1.1mm]
K_\mathrm{sym}=&-\frac{10}{9}\frac{\hbar^2}{2m}\frac{\tau_0}{\rho_0}
-B_0^{\rho_0}\left[\frac{5\xi^2}{3}G_0(\xi)
+\frac{4\xi^3}{3}G_0'(\xi)+\frac{\xi^4}{6}G_0''(\xi)\right]\rho_0
+\frac{25}{9}\left(A_1^{\rho_0}+A_2^{\rho_0}\right)\tau_0  \nonumber \\[-1.1mm]
&+\frac{1}{9}\left(B_1^{\rho_0}-B_2^{\rho_0}\right)
\left[25G_1(\xi)-14\xi^2G_2(\xi)+20\xi G_1'(\xi)-6\xi^3G_2'(\xi)
+\frac{5\xi^2}{2}G_1''(\xi)-\frac{\xi^4}{2}G_2''(\xi)\right] \tau_0\nonumber
 \\[-1.1mm]
&+\frac{25}{3}\left(B_1^{\rho_1}-B_2^{\rho_1}\right)
\left[H_0(\xi)
+\frac{4\xi}{5}H_0'(\xi)+\frac{\xi^2}{10}H_0''(\xi)
-\frac{7\xi^2}{15}H_1(\xi)-\frac{\xi^3}{5}H_1'(\xi)
-\frac{\xi^4}{60}H_1''(\xi)\right] \tau_0 \nonumber \\[-1.1mm]
&+B_0^{\rho_1}\left[4\xi H_0'(\xi)+\xi^2H_0''(\xi)\right] \rho_0
+\frac{25}{3}\left(A_1^{\rho_1}+A_2^{\rho_1}\right)\tau_0 \,.  \\[-10mm]  \nonumber
\end{align}

In the present study, we present results for two preliminary sets of
$a=0.8$\,fm parameters. First, by using conditions
$T_2^{(i)}=-T_1^{(i)}$, we obtained parameters REG2a.130531 that
correspond to a local finite-range momentum-independent
potential~\cite{Dobacweski12}. Second, by releasing these
constraints, we obtained parameters REG2b.130531. On the one hand,
both sets correspond to the same values of
$\rho_\mathrm{sat}=0.16$\,fm, $E/A=-16$, $K_\infty=230$, and $J=32$.
On the other hand, they correspond to different values of $L=100.2$
and 58, $K_\mathrm{sym}=83.26$ and $-$175, and $m^*/m=0.38$ and 0.41,
respectively (all energies are in MeV). Values of
parameters are listed in Table 1. In both cases, saturation in
symmetric matter is
obtained through the interplay between the $T_0^{(i)}$ attractive
and $T_1^{(i)}$ and $T_2^{(i)}$ repulsive terms.

In Fig.~1 we show equations of state (EOS) for symmetric, neutron,
polarized symmetric and polarized neutron matter for the effective
interactions Skyrme SV~\cite{Beiner75} (zero-range density-independent),
Gogny D1N~\cite{Chappert08}
(finite-range density-dependent) and the two regularized $\delta$ interactions.
The Skyrme SV interaction is the only two-body density-independant interaction
commonly used nowadays while D1N is a succesful finite-range density-dependent
interaction.
The regularized $\delta$ interactions have effective masses
that are too low at saturation density but a more realistic incompressibility,
similar to that obtained with D1N. Different dependences on momenta
and densities of the four EOS make their behavior in neutron,
polarized symmetric, and polarized neutron matter very different, although
symmetric matter remains the ground state upto very high densities.

\vspace*{0.2cm}
\noindent
\centerline{\includegraphics[width=0.9\textwidth]{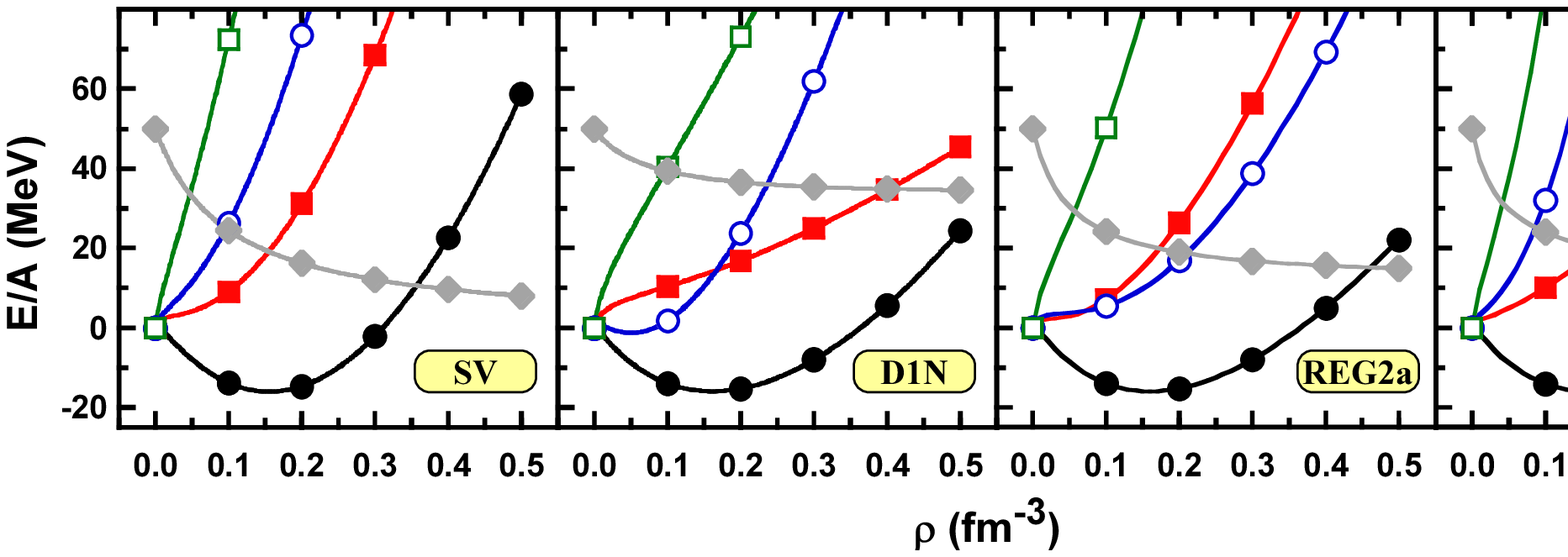}}
{\small{\bf Figure 1.}
Equation of state of the symmetric (full circles), neutron (full squares),
polarized symmetric (open circles), and polarized neutron (open squares)
infinite matter (left scale). The effective mass (right scale) is shown with
diamonds.}%

\newlength{\lesswid}
\setlength{\lesswid}{0.2cm}
\noindent
\begin{minipage}[t]{0.59\textwidth}
\noindent
{\small{\bf Table 1.} Values of coupling constants $T_k^{(i)}$
that define the $a=0.8$\,fm parametrizations REG2a.130531
(top, $T_2^{(i)}=-T_1^{(i)}$) and REG2b.130531
(bottom) of the
regularized $\delta$ interaction (\protect\ref{eq:bdr}),
in units of MeV\,fm$^{3}$ ($k=0$) and MeV\,fm$^{5}$ ($k=1,2$).
Standard zero-range spin-orbit forces with
$W_0=209.30$ and 168.35~MeV\,fm$^{5}$, respectively, were used.}%
\begin{center}
\begin{tabular}{r|r@{\hspace*{\lesswid}}r@{\hspace*{\lesswid}}r@{\hspace*{\lesswid}}r}
\hline
$T_k^{(i)}$& $i=1$      & $i=2$          & $i=3$          & $i=4$           \\
\hline
 $k=0$ &  $-$968.645    &    1645.515    & $-$1400.680    &  $-$451.892     \\
 $k=1$ &  $-$653.038    &    1349.673    & $-$2011.063    &    1692.524     \\
\hline
 $k=0$ & $-$12250.143  &    7277.075  &$-$6952.679  &   10744.723  \\
 $k=1$ &  $-$1149.333  &    1594.666  &$-$2342.666  &    2413.333  \\
 $k=2$ &     3184.584  &  $-$513.696  &   4681.530  & $-$5127.553  \\
\hline
\end{tabular}
\end{center}
\end{minipage}\hspace{0.03\textwidth}%
\begin{minipage}[t]{0.38\textwidth}
\vspace*{-0.2cm}
\begin{center}
\includegraphics[width=\textwidth]{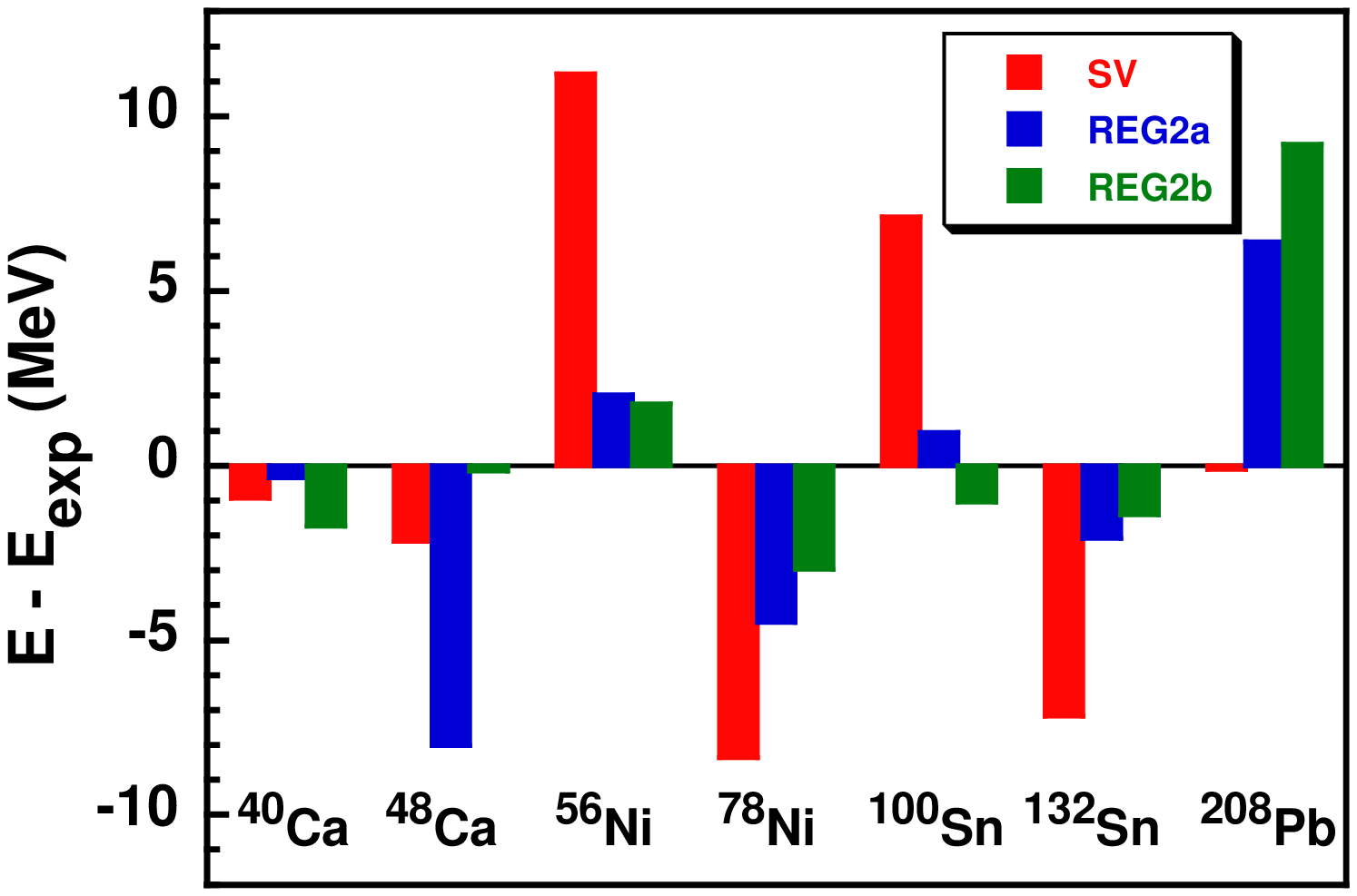}%
\end{center}
\noindent
{\small{\bf Figure 2.} Deviations of ground-state energies of doubly magic
nuclei from experimental data.}%
\end{minipage}%
\vspace*{0.5cm}

To calculate finite nuclei, in the HFODD (v2.64p)
solver~\cite{Schunck12} we implemented self-consistent
solutions for the regularized $\delta$ interaction (\ref{eq:bdr}).
Masses of doubly magic nuclei are shown in Fig.~2.
We see that REG2a.130531 and, even more, REG2b.130531 represent an
improvement compared to the Skyrme SV interaction.

The new effective interaction presented in this work
seems to be promising.
Although it
does not contain any density-dependent term, it reproduces all
empirical properties of the saturation point but the isoscalar effective mass,
which is rather low. To our knowledge, this
has never been achieved with any other density independent two-body
interaction.
Since this interaction has a finite range, it
can be used in any calculation at the mean field level or beyond,
without the need to introduce any additional momentum cut-off.

This work has been supported in part by the Academy of Finland and
University of Jyv\"askyl\"a within the FIDIPRO programme and by the
Polish French agreement COPIN-IN2P3 Project no.\ 11-143. We
acknowledge the CSC-IT Center for Science Ltd, Finland
and Centre de Calcul CC-IN2P3 (IN2P3, CNRS Villeurbanne, France)
for the allocation of computational resources.

\bibliographystyle{unsrt}

%
%
\end{document}